\documentclass[twocolumn,showpacs,preprintnumbers,amsmath,amssymb, prb, reprint]{revtex4-1}
\usepackage[english]{babel}
\usepackage{amsmath}
\usepackage{stmaryrd}
\usepackage{txfonts}
\usepackage{amssymb}
\usepackage{mathrsfs}
\usepackage{graphicx}
\usepackage{dcolumn}
\usepackage{bm}
\usepackage{epsfig}
\usepackage{color}
\usepackage[colorlinks=true, linkcolor=blue, citecolor=blue, urlcolor=blue]{hyperref} 

\begin{document}
\preprint{\href{http://link.aps.org/doi/10.1103/PhysRevB.91.224407}{S.-Z. Lin, A. Saxena and C. D. Batista, Phys. Rev. B {\bf 91}, 224407 (2015).}}

\title{Skyrmion fractionalization and merons in chiral magnets with easy-plane anisotropy}

\author{Shi-Zeng Lin}
\affiliation{Theoretical Division, Los Alamos National Laboratory, Los Alamos, New Mexico
87545, USA}
\author{Avadh Saxena}
\affiliation{Theoretical Division, Los Alamos National Laboratory, Los Alamos, New Mexico
87545, USA}
\author{Cristian D. Batista}
\affiliation{Theoretical Division, Los Alamos National Laboratory, Los Alamos, New Mexico
87545, USA}

\begin{abstract}
We study the equilibrium phase diagram of ultrathin chiral magnets with  easy-plane anisotropy $A$. The
vast  triangular skyrmion lattice phase that is stabilized by an external magnetic field evolves continuously as a function 
of increasing $A$ into a regime in which nearest-neighbor skyrmions start overlapping with each other. This overlap leads to
a continuous  reduction of the skyrmion number from its quantized value $Q=1$ and to the emergence of antivortices at the center of the triangles formed 
by nearest-neighbor skyrmions.  The antivortices also carry a small ``skyrmion number'' $Q_A \ll 1$ that grows as a function of increasing $A$.
The system undergoes a first order phase transition into a square vortex-antivortex lattice at a critical value of $A$. Finally, a canted ferromagnetic state
becomes stable through another first order transition for a large enough anisotropy $A$. Interestingly enough, this  first order transition is accompanied by
{\it metastable} meron solutions. 
\end{abstract}
\date{\today}
\maketitle

\section{Introduction}

Emergent stable topological structures are ubiquitous in condensed matter, high energy physics and materials science, as well as in atomic, molecular and optical physics.  Prominent examples include domain walls, vortices, dislocations, disclinations and skyrmions.   Skyrmions are vector field (e.g. spin) textures \cite{Skyrme61}, which wrap a sphere once, and can be viewed as emergent mesoscale particles. These textures have been observed in quantum Hall states of  two-dimensional electron gases, chiral nematic liquid crystals, and Bose-Einstein condensates, among others. 

\begin{figure}[b]
\psfig{figure=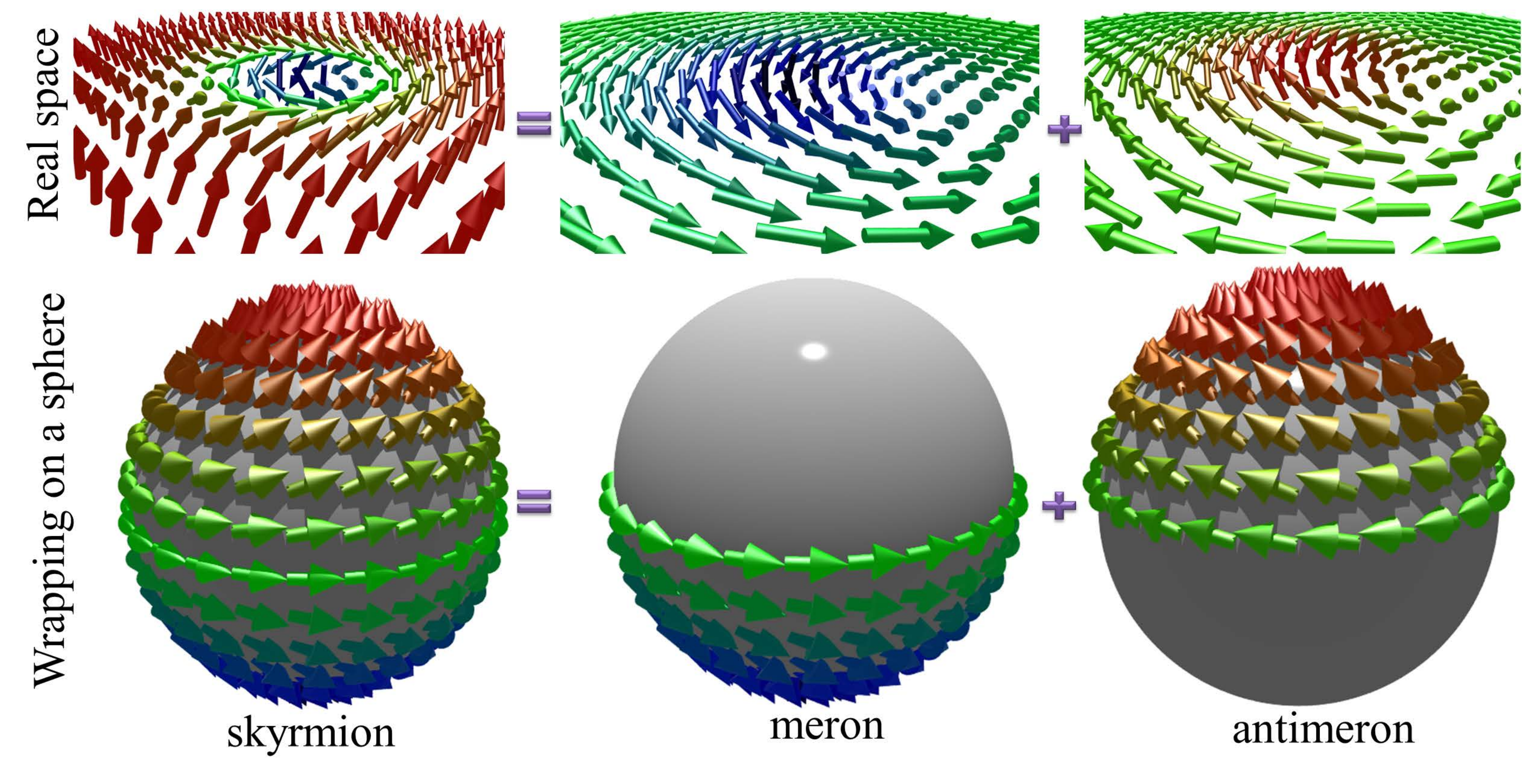,width=\columnwidth}
\caption{(color online) Upper panel: Dissociation of a skyrmion into a meron and antimeron pair. Lower panel: A skyrmion covers the sphere, whereas a meron  covers only half the sphere, meaning that the topological charge of a skyrmion is $|Q|=1$ and that of a meron is $|Q|=1/2$.  The regions where the spins are parallel  and antiparallel to the applied field are indicated with red and blue colors respectively.  Spins are parallel to the plane in the green (equatorial) region.
} \label{f1}
\end{figure}

Skyrmion crystals~\cite{Bogdanov89,Rosler2006} have been recently discovered in chiral magnets ~\cite{Muhlbauer2009, Yu2010a,Yu2011,Seki2012,Adams2012} and they have triggered enormous interest because of their significant potential for spintronics applications~\cite{Fert2013}. Each skyrmion  consists of a spin configuration that wraps the sphere once over a length scale that ranges from 10 to 100 nm  [see Fig.~\ref{f1}(a)]. This property is quantified by the skyrmion topological charge $Q=\frac{1}{4\pi}\int_{\mathrm{Sk}} dr^2 \mathbf{n}\cdot( \partial_x \mathbf{n}\times \partial_y \mathbf{n})=\pm1$ [$\mathbf{n}(\mathbf{r})$ is a unit vector describing the direction of the spin with spatial coordinates $\mathbf{r}$]. Because of the large Hund's coupling between conduction electrons and spins in metals, the spin of the conduction electrons is fully aligned with the magnetic  moments that form the skyrmion texture. A conduction electron picks up a  Berry phase of $2\pi$  when it moves around a skyrmion. This is equivalent to 
the Berry phase induced by one flux quantum $\Phi_0=hc/e$, i.e. ,  the effective flux produced by a skyrmion is $\Phi= Q \Phi_0$. The equivalent  magnetic field is enormous: 100 T for a skyrmion of  radial size 10 nm.  This phenomenon leads to the so-called  topological Hall effect ~\cite{Neubauer2009,NagaosaRMP2010,Zang11}. Indeed, most of the excitement generated by the recent discovery of mesoscale skyrmions in chiral magnets arises from their transport properties. Skyrmions  can be driven by  spin polarized currents in metals~\cite{Jonietz2010,Yu2012,Schulz2012} or by  magnon currents  in insulators ~\cite{Kong2013,Lin2014PRL,Mochizuki2014}. A very attractive feature is  that the currents required to move  skyrmions in chiral magnets  are $10^5$ to $10^6$ times smaller than the depinning currents of  magnetic domain walls ~\cite{Jonietz2010,Yu2012,Schulz2012}, making skyrmions  prime candidates for novel information storage devices with substantially reduced power consumption ~\cite{Fert2013}.

Fractionalization of a topological object is another  active topic in condensed-matter and high energy physics. By splitting  a  skyrmion into two halves, we obtain a  meron and an antimeron (see Fig. ~\ref{f1}). The meron wraps only half of the sphere, implying that its topological charge is $Q=\pm1/2$. Similar to skyrmions, merons also have many different physical realizations. For instance, merons become stable in small magnetic disks because of a strong demagnetization effect:   spins align along the disk boundary in the outer region, but they are normal to the surface near the center of the disk ~\cite{Chien2007}. Like in the case of skyrmions, meron crystals have been predicted to exist in quantum Hall systems ~\cite{Brey1996}, as well as  in bilayer graphene ~\cite{Coter2010}.

The main purpose of this work is to study the evolution of a skyrmion crystal as a function of increasing easy-plane anisotropy $A$ in a two-dimensional system. It is clear that the easy-plane anisotropy penalizes the non-coplanar nature of the skyrmion texture. As we will see below, this effect leads to a continuous reduction of the skyrmion number from its quantized
$|Q|=1$ value when the skyrmions start overlapping with each other. 
While the   skyrmion charge $|Q|$ is no longer equal to one, the winding number is still an integer, i.e., the resulting 
spin configuration can be regarded as a crystal of vortices/antivortices with a fractional  skyrmion charge $|Q_{v,a}| <1 $. The second consequence of the overlap between skyrmions is
the emergence of antivortices at the centers of the triangles formed by three nearest-neighbor skyrmions. These antivortices also carry a small ``skyrmion number'' $|Q_a| \ll 1$ that grows as a function of increasing $A$ and eventually triggers a first order transition into a square vortex-antivortex crystal. In this new phase, the vortices and antivortices are close to becoming
merons in the sense that their skyrmion charge is close to 1/2: $|Q_{v,a}| \simeq 1/2$. Vortices and antivortices are indeed expected to become merons and anti-merons ($|Q_v|=|Q_a|=1/2$) when the separation between them becomes larger than the diameter of their non-coplanar cores. Unfortunately,
a first order transition into a canted ferromagnet takes place right before this condition is fulfilled. However, this  transition is accompanied
by {\it metastable} meron states in which a subset of vortices  remain well separated from the other vortices and antivortices and can therefore be regarded as merons.

We note that  the effect of easy-plane anisotropy on the skyrmion lattice phase has been considered in a previous work \cite{Banerjee2014}. The authors used a  variational treatment that
only contains the triangular skyrmion crystal and the canted ferromagnet as the possible candidates. In contrast, the  Langevin dynamics approach that we use in the present work
is completely unbiased.  The phase diagram obtained with the variational treatment differs qualitatively from 
the one that we are reporting here simply because the above described square vortex-antivortex crystal is outside the variational space of Ref.~\onlinecite{Banerjee2014}. 
Nevertheless, the variational calculation based on a square unit cell indicates a tendency to form such a square lattice. 
Our unbiased calculation also unveils interesting metastable meron states that cannot be detected with the variational treatment.

Finally, we note that the same model has been studied in Ref.~\onlinecite{Wilson2014} for the case of films with finite thickness of the order of the wave-length of the spiral.
In this case, the easy-plane anisotropy stabilizes the conical phase relative to the skyrmion crystal phase for the model under consideration. Consequently, 
the above described transition between the triangular skyrmion crystal and the square vortex-antivortex lattice was not reported in Ref.~\onlinecite{Wilson2014} . 

 \begin{figure*}[t]
\psfig{figure=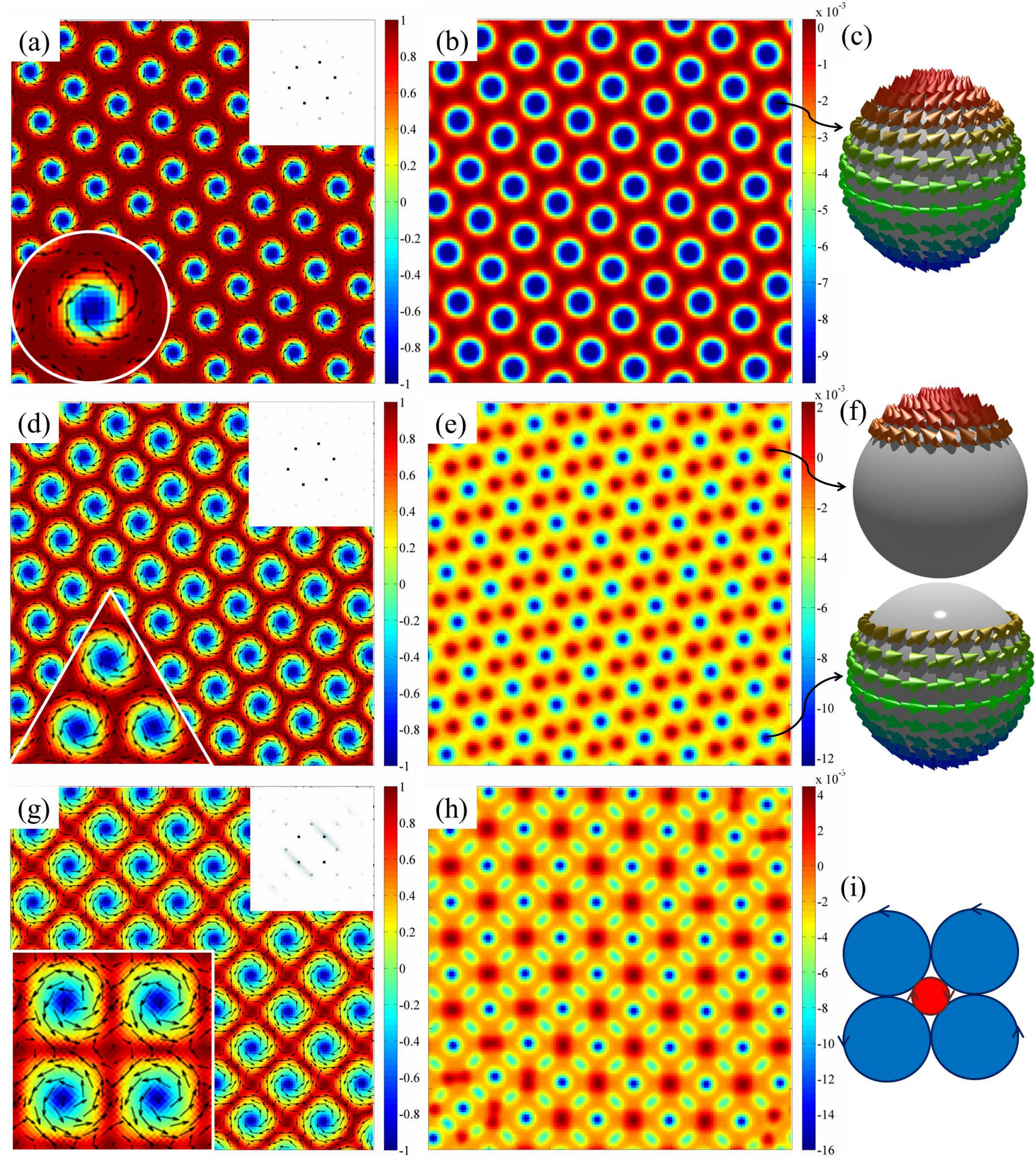,width=16.2cm}
\caption{(color online) Evolution of the spin texture as a function of increasing easy-plane anisotropy. The left column is the spin texture. The color denotes the spin component along the $z$ direction, while the arrows represent the in-plane spin component. The middle column illustrates the topological charge density $q(\mathbf{r})$. From top to bottom, $A=0.075 D^2/J_{\mathrm{ex}}$, $A=0.825 D^2/J_{\mathrm{ex}}$, and $A=1.275 D^2/J_{\mathrm{ex}}$. There is a transition from a triangular to a square lattice with increasing $A$. The upper insets in (a, d, g) are the Fourier transform of $n_z$ after subtracting the uniform component, while the lower insets are enlarged views. (c) and (f) show the corresponding wrapping of spin textures on the surface of a sphere. (i) illustrates why the chirality of the meron at the center is opposite to the surrounding merons. Here $B_z=0.7 D^2/J_{\mathrm{ex}}$ and $L_x\times L_y=60 \times 60$ $(J_{\mathrm{ex}}a/D)^2$. The results are qualitatively the same for other magnetic fields, see Fig. \ref{f6}.
} \label{f2}
\end{figure*}

\section{Model}

We consider the two-dimensional Hamiltonian 
\begin{equation}\label{eq1}
\mathcal{H}=\int d\mathbf{r}^2 \left[\frac{J_{\rm{ex}}}{2}(\nabla \mathbf{n})^2+\frac{D}{a}\mathbf{n}\cdot(\nabla\times \mathbf{n})-\frac{B_z}{a^2} n_z+\frac{A}{a^2} n_z^2 \right],
\end{equation}
that describes a single layer of a chiral magnet with easy-plane anisotropy ~\cite{Bogdanov89,Bogdanov94,Rosler2006,Han10,Rossler2011}. Here  $a$ is the lattice constant, $J_{\rm{ex}}$ is the ferromagnetic exchange and $D$ is the Dzyaloshinskii-Moriya (DM) interaction \cite{Dzyaloshinsky1958,Moriya60,Moriya60b} generally present in magnets without inversion symmetry. $B_z$ is the external magnetic field  normal to the film and $A>0$ is the easy-plane anisotropy. 
Equation \eqref{eq1} describes satisfactorily the experimental phase diagram \cite{Yu2010a}. To find the different stable phases as a function of increasing easy-plane anisotropy, $A$, we simulate the dynamics of the spin field $\mathbf{n}({\bf r})$ by numerically solving the  Landau-Lifshitz-Gilbert equation with a noise term \cite{Lin2014PRL}
\begin{equation}\label{eq2}
	{\partial _t}{\bf{n}} =- \gamma {\bf{n}} \times ({{\bf{B}}_{\rm{eff}}+\tilde{\mathbf{B}}}) + \alpha \mathbf{n}\times  {\partial _t}{\bf{n}}+\frac{\hbar\gamma}{2e}(\mathbf{J}\cdot\nabla)\mathbf{n}.
\end{equation}
The effective field is $\mathbf{B}_{\rm{eff}}\equiv-\delta \mathcal{H}/\delta {\bf{n}}={J_{\mathrm{ex}}}{\nabla ^2}{\mathbf{n}} - 2D\nabla  \times {\mathbf{n}} + {\mathbf{B}}- 2A n_z\hat{z}$, $\gamma=1/(\hbar s)$, and $s$ is the magnitude of local spins. The unit vector $\hat{z}$ is parallel to the $z$ direction, $\alpha$ is the Gilbert damping coefficient and $\mathbf{J}$ is the external spin polarized current. $\tilde{\mathbf{B}}$ is the fluctuating magnetic field that satisfies the fluctuation-dissipation theorem: $\langle \tilde{\mathbf{B}} \rangle =0$ and
\begin{equation}\label{eq3}
	\langle \tilde{B}_{\mu}(\mathbf{r}, t)\tilde{B}_{\nu}(\mathbf{r}', t') \rangle=\frac{2 k_B T\alpha a^2}{ \gamma} \delta_{\mu,\nu}\delta(\mathbf{r}'-\mathbf{r})\delta(t'-t),
\end{equation}
where $T$ is the temperature, $k_B$ is the Boltzmann constant and $\mu, \nu=x,\ y,\ z$. The $T=0$ state is obtained by slowly annealing the system from the paramagnetic state without an external current ($J=0$).  

\section{Results}

\begin{figure*}[t]
\psfig{figure=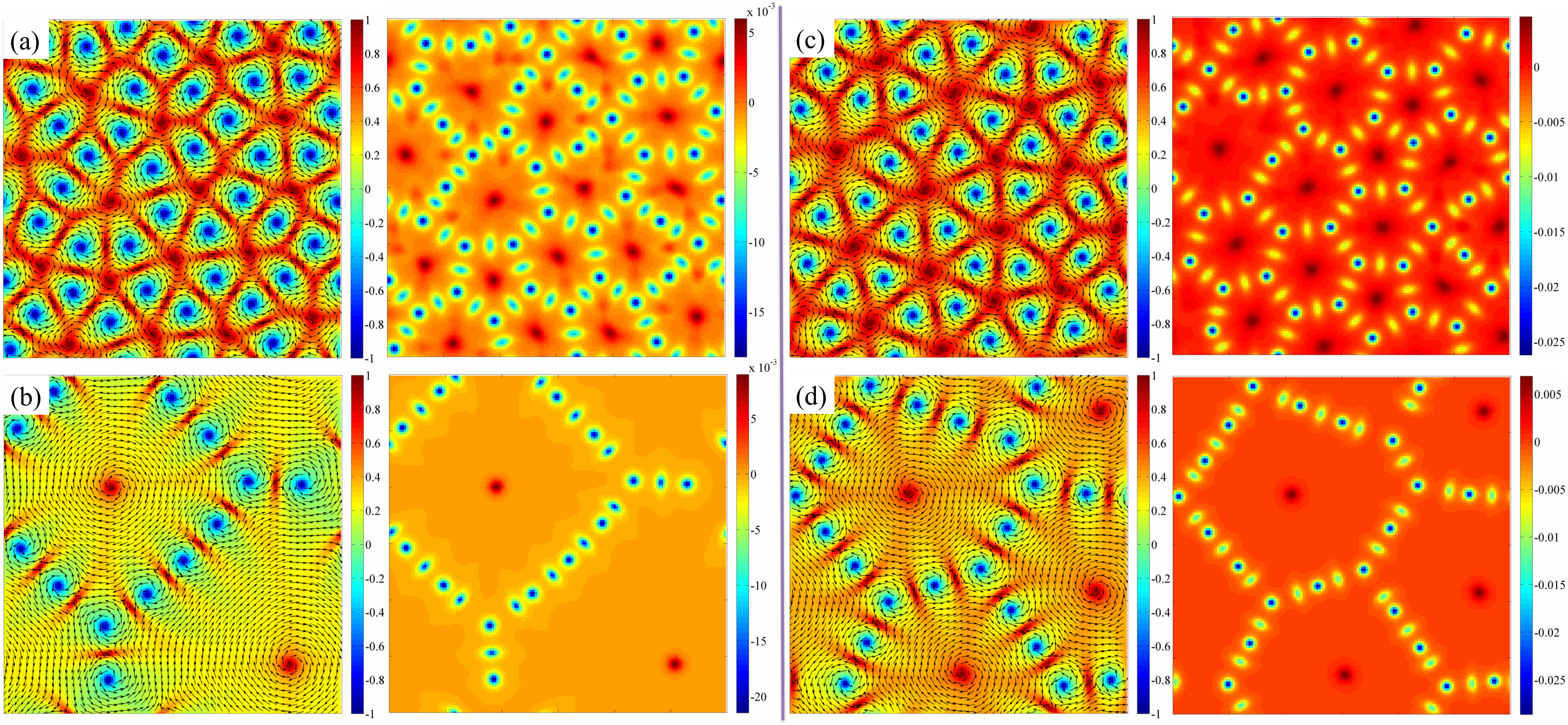, width=16.2cm}
\caption{(color online) Metastable states that accompany the first order transition between the square lattice depicted in Figs.~\ref{f2}(g) and \ref{fs1}(e)  and the canted ferromagnetic state. 
These states appear after the annealing from the paramagnetic state near the  boundary between both phases. 
Panels (a) and (b) correspond to  $A=1.5 D^2/J_{\mathrm{ex}}$ and $A=1.875 D^2/J_{\mathrm{ex}}$ respectively for $B_z=0.7D^2/J_{\rm ex}$.  
Panels (c) and (d) correspond to  $A=1.65 D^2/J_{\mathrm{ex}}$ and $A=1.95 D^2/J_{\mathrm{ex}}$ respectively for $B_z=1.2 D^2/J_{\rm ex}$.
 Meron (blue) loops containing an anitmeron (red) are clearly visible.  Show are the spin configuration and the corresponding skyrmion topological charge density profile.} 
\label{f3}
\end{figure*}

The evolution of the low temperature spin texture as a function of $A$ is depicted in the left column of Fig.~\ref{f2}. The  second column of Fig.~\ref{f2}  shows the corresponding skyrmion topological charge density $q(\mathbf{r})=\mathbf{n}\cdot( \partial_x \mathbf{n}\times \partial_y \mathbf{n})/(4\pi)$.  For a small $A$, we recover the expected triangular skyrmion crystal.  The spins are anti-aligned with the magnetic field at the center of each skyrmion, while they remain parallel to the field in the background region between skyrmions. As the easy-plane anisotropy increases, the skyrmion size also increases because of a natural expansion of the equatorial region  favored by $A$.  Skyrmions start overlapping with each other as a consequence of this expansion, implying that the spins no longer wrap the full sphere, $|Q|<1$, because the skyrmion size becomes larger than the separation between neighboring skyrmions.  The growth of the equatorial regions of three skyrmions that close a triangular loop [see Fig.~\ref{f2}(d)], or a square loop [see Fig.~\ref{f2}(g)], generates a vortex with an opposite winding number (antivortex)  at the center of the loop. This antivortex is actually the precursor of an antimeron because the spins in the core region remain canted along the field direction but $|Q_a| <1/2$. We also note that the continuous suppression of the skyrmion charge $|Q|$ of the blue ``particles" in Figs.~\ref{f2} and  \ref{f3} implies that the former skyrmions ($Q=1$) of Fig.~\ref{f2}(a) evolve continuously into the vortices with $|Q_v| >1/2$ that are shown in Figs.~\ref{f2}(d) and \ref{f2}(g). 
The vortices and antivortices become merons and anti-merons ($|Q|=1/2$) when the separation between them becomes bigger than the diameter of the non-coplanar cores. This clear separation only occurs for the metastable states shown in Fig.~\ref{f3}. 

 \begin{figure*}[t]
\psfig{figure=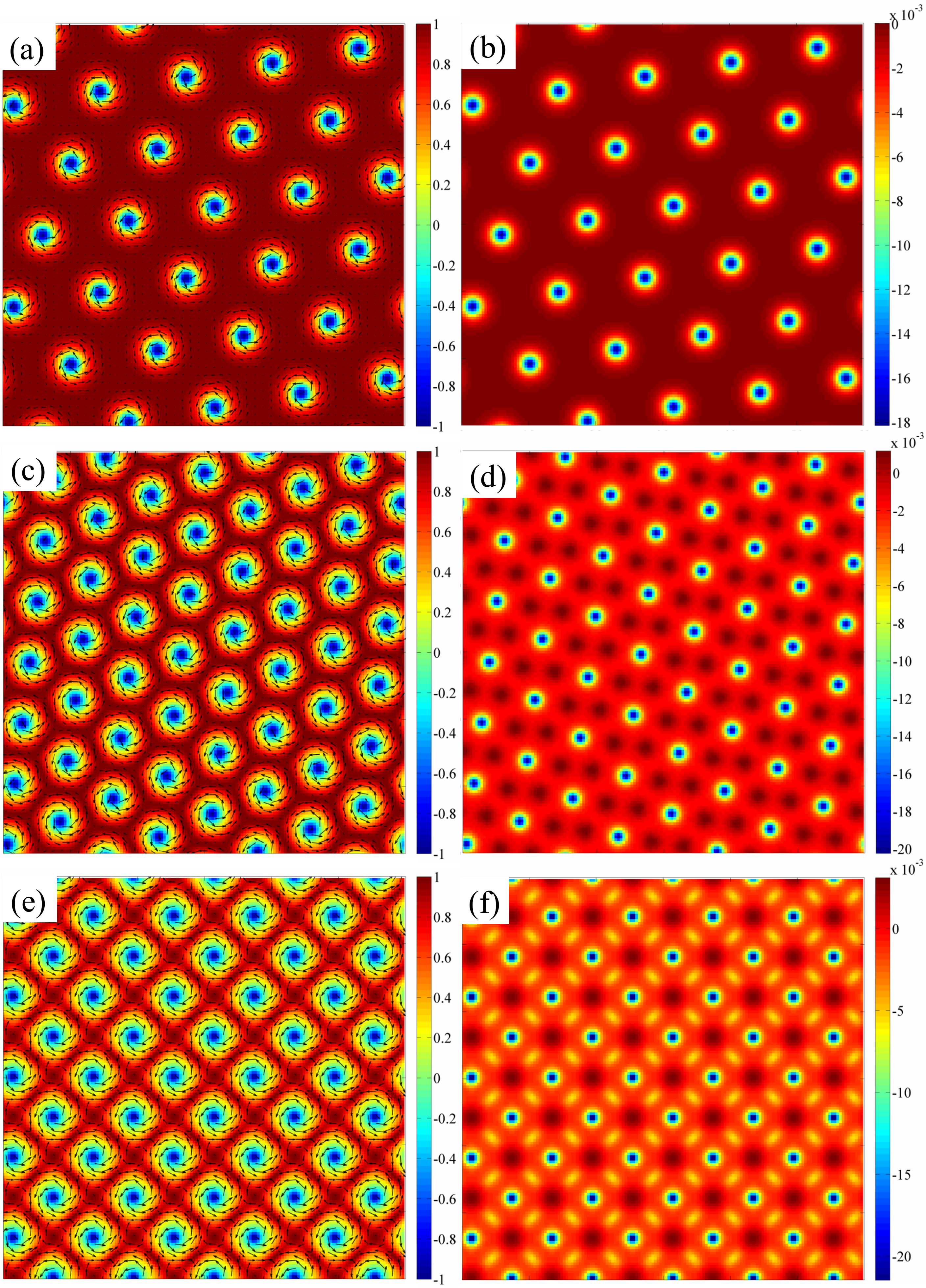,width=13.2cm}
\caption{(color online) Evolution of the spin texture as a function of increasing easy-plane anisotropy. The left column is the spin texture. The color denotes the spin component along the $z$ direction, while the arrows represent the in-plane spin component. The right column illustrates the topological charge density $q(\mathbf{r})$. From top to bottom, $A=0.525 D^2/J_{\mathrm{ex}}$, $A=0.975 D^2/J_{\mathrm{ex}}$, and $A=1.425 D^2/J_{\mathrm{ex}}$. There is a transition from a triangular to a square lattice with increasing $A$. Here $B_z=1.2 D^2/J_{\mathrm{ex}}$ and $L_x\times L_y=60 \times 60$ $(J_{\mathrm{ex}}a/D)^2$. Upon increasing $A$ from 0, the system evolves in the following sequence: spin polarized ferromagnetic state, triangular lattice of skyrmions, square lattice of fractional skyrmions, spin canted ferromagnetic state.} 
\label{fs1}
\end{figure*}

As it is shown in the different panels of Fig.~\ref{f2}, the vortices  (blue particles) and antivortices (red particles) form two different crystals. For $A=0.825 D^2/J_{\mathrm{ex}}$, the crystal consists of a triangular lattice with  blue particles in one sublattice and red particles in the other two [see Figs.~\ref{f2} (d) and \ref{f2}(e)]. The second crystal becomes stable for $1.275 D^2/J_{\mathrm{ex}}<A<1.425 D^2/J_{\mathrm{ex}}$ and consists of two inter-penetrating square lattices of blue and red particles [see Figs.~\ref{f2}(g) and \ref{f2}(h)] . If we think of the blue and red particles as disks of different radii, these two crystals are the only compact packings~\cite{Kennedy04} that satisfy the following rule: two small disks cannot be tangent to each other. This rule arises from an energetic consideration: spins near the point of contact of two disks of the same type (chirality) have to be parallel to the hard axis.  The triangular and  square crystals of Fig.~\ref{f2}(d) and ~\ref{f2}(g) become compact packings when the ratio between the radii of the red and blue particles is  equal to $r_r/r_b=2 \sqrt{3}/3 -1$ and $r_r/r_b=\sqrt{2}-1$, respectively ~\cite{Kennedy04}. Because $r_r/r_b$ increases with $A$, it is natural to expect a transition from the triangular [Fig.~\ref{f2} (d)] to the square [Fig.~\ref{f2} (g)]  interpenetrating crystals \footnote{The transition from a triangular to square lattice was studied in Ref. \onlinecite{Yi09} with a similar Hamiltonian.}. Figure~\ref{fs1}  shows a similar  evolution of the spin texture as a function of increasing easy-plane anisotropy  for $B_z=1.2 D^2/J_{\mathrm{ex}}$.

As the easy-plane anisotropy is further increased, we expect that the particles should separate from each other because of the continuous growth of equatorial region of the skyrmion [see Fig.~\ref{f1}]. However,
a first order transition into a  canted ferromagnetic state precludes the possibility of observing this behavior in equilibrium. Nevertheless, the expected behavior is observed 
in the metastable states that accompany this first order transition. As it is shown in  Fig.~\ref{f3}, the background of these states consists of a coplanar spin ordering in the $xy$-plane. 
The former skyrmions or blue particles are now loops of vortices and antivortices with negative scalar spin chirality of skyrmion charge. 
At the center of the vortex-antivortex loops (blue particles), there is always a $Q_a=1/2$ antimeron  (red particles), at it is shown in Fig.~\ref{f3}.

The effective interaction between merons decreases with their separation rendering the meron spin texture softer relative to the insertion of topological defects, such as dislocations and disclinations. Indeed, these topological defects are already present in the metastable meron-antimeron configurations shown in Fig.~\ref{f3}.  While the metastable states preserve a net total skyrmion charge or scalar chirality, the total winding number  must be zero because the spins are mostly parallel to the plane, except for the isolated regions occupied by the vortex cores, and the boundary conditions are periodic. This implies that the number of vortices is equal to the number of antivortices. We note however that the distribution of vortices and antivortices is far from being uniform.   For instance, the state depicted in Fig.~\ref{f3}(b) includes a subset of antivortices (red circular regions) that remain distant from the rest of the vortices and antivortices. The isolated character of these antivortices leads to a half-quantized skyrmion number $Q=1/2$ that makes them antimerons. The same is not true for the chains of alternating vortices and antivortices that surround each of these merons  because their cores overlap with each other [Fig.~\ref{f3}(b)].
Figures~\ref{f3}(c) and (d)  show  similar  metastable states for a higher  magnetic field value, $B_z=1.2 D^2/J_{\mathrm{ex}}$, indicating that the basic structure of these states and their local stability are rather independent of $B_z$.

\begin{figure}[t]
 	\psfig{figure=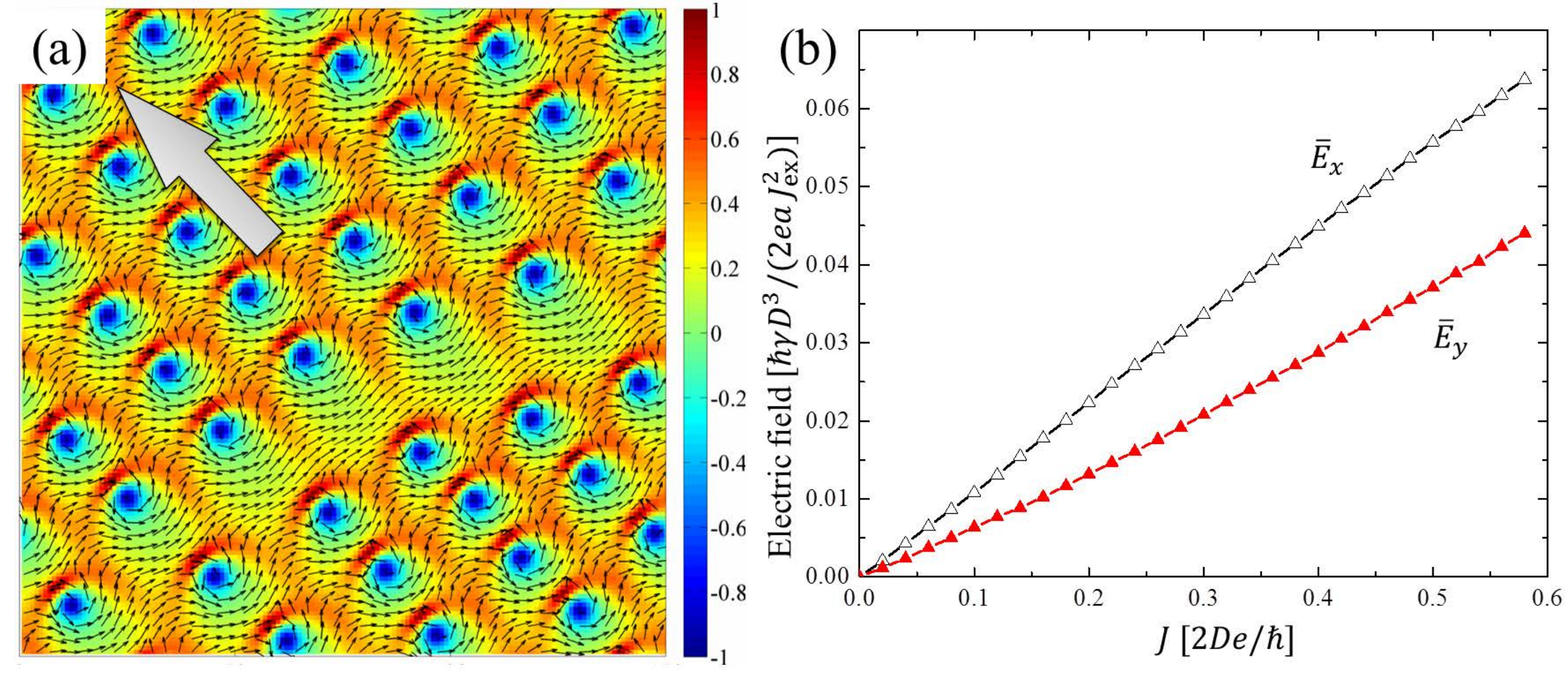, width=\columnwidth}
 	\caption{(color online) (a) Snapshot of meron flow at $J_x=0.6De/\hbar$ and $J_y=0$ with $A=1.5 D^2/J_{\mathrm{ex}}$ and $\alpha=1.0$. The arrow represents the direction of motion. (b) The averaged electric field versus current. } \label{f4}
 \end{figure}

Although the states depicted in Fig.~\ref{f3} are just metastable, the vortex-antivortex pairs generated after the annealing process  remain locally stable even after  injecting a finite current into the system.  In contrast, the isolated antimerons move in the opposite direction (they have opposite scalar chirality) and  get annihilated after colliding with an meron. The steady state, cf. Figs. \ref{f3} (a) and (c), is a uniform flow of vortex-antivortex pairs. As it is shown in Fig.~\ref{f4}~(a) these pairs become  dynamically ordered along the direction of motion. Figure~\ref{f4}~(b) shows the averaged electric field, $\mathbf{E}={\hbar}\mathbf{n}\cdot(\nabla\mathbf{n}\times\partial_t\mathbf{n})/({2e})$, generated by this motion as a function of the applied electric current.

\section{Phase Diagram}
 
\begin{figure}[b]
\psfig{figure=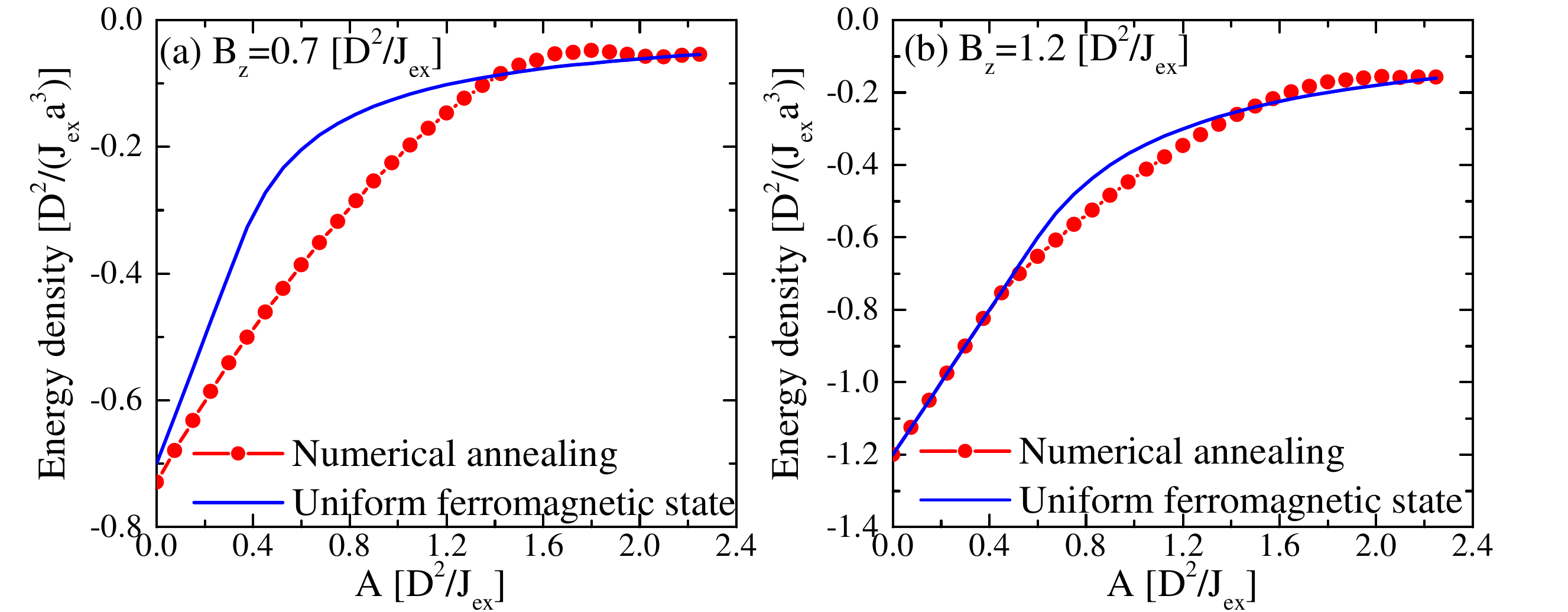, width=\columnwidth}
\caption{(color online) (a) Comparison of energy obtained by numerical annealing and that of the canted ferromagnetic state at (a) $B_z=0.7 D^2/J_{\mathrm{ex}}$ and (b) $B_z=1.2 D^2/J_{\mathrm{ex}}$.} 
\label{f5}
\end{figure}

To construct an anisotropy-applied magnetic field ($A$-$B_z$) phase diagram, we compare the energy of the states obtained by annealing to that of the normal spiral state and the uniform  (canted) ferromagnetic state. The uniform ferromagnetic state can either be  the spin polarized ferromagnetic state along the magnetic field direction ($n_z=1$) or the spin canted ferromagnetic state with $n_z<1$. There is a phase transition line separating these two states because only the latter one breaks the $U(1)$ symmetry of global spin rotation along the $z$-axis. The energy obtained from numerical annealing is compared against the analytical expression  for the uniform ferromagnetic state, as it is shown in Fig. \ref{f5}. State of the art numerical methods are not adequate for finding the optimal state of equilibrium near first order phase transitions. We note in passing that this statement  also applies to real physical systems: they  exhibit hysteresis near first order phase transitions because they get trapped in metastable states. For this reason,  it is still relevant to understand the nature of these states. As anticipated in the previous section, our numerical annealing gets trapped in the metastable states depicted in Fig.~\ref{f3} near the first order phase boundary between the square lattice depicted in Fig.~\ref{f2}(g) and the spin canted ferromagnetic state. While this recurrent problem limits the accuracy of the obtained phase transition lines, the unbiased nature of our  numerical method enables the possibility of finding new phases that have been missed by the variational treatments.

\begin{figure}[t]
 \psfig{figure=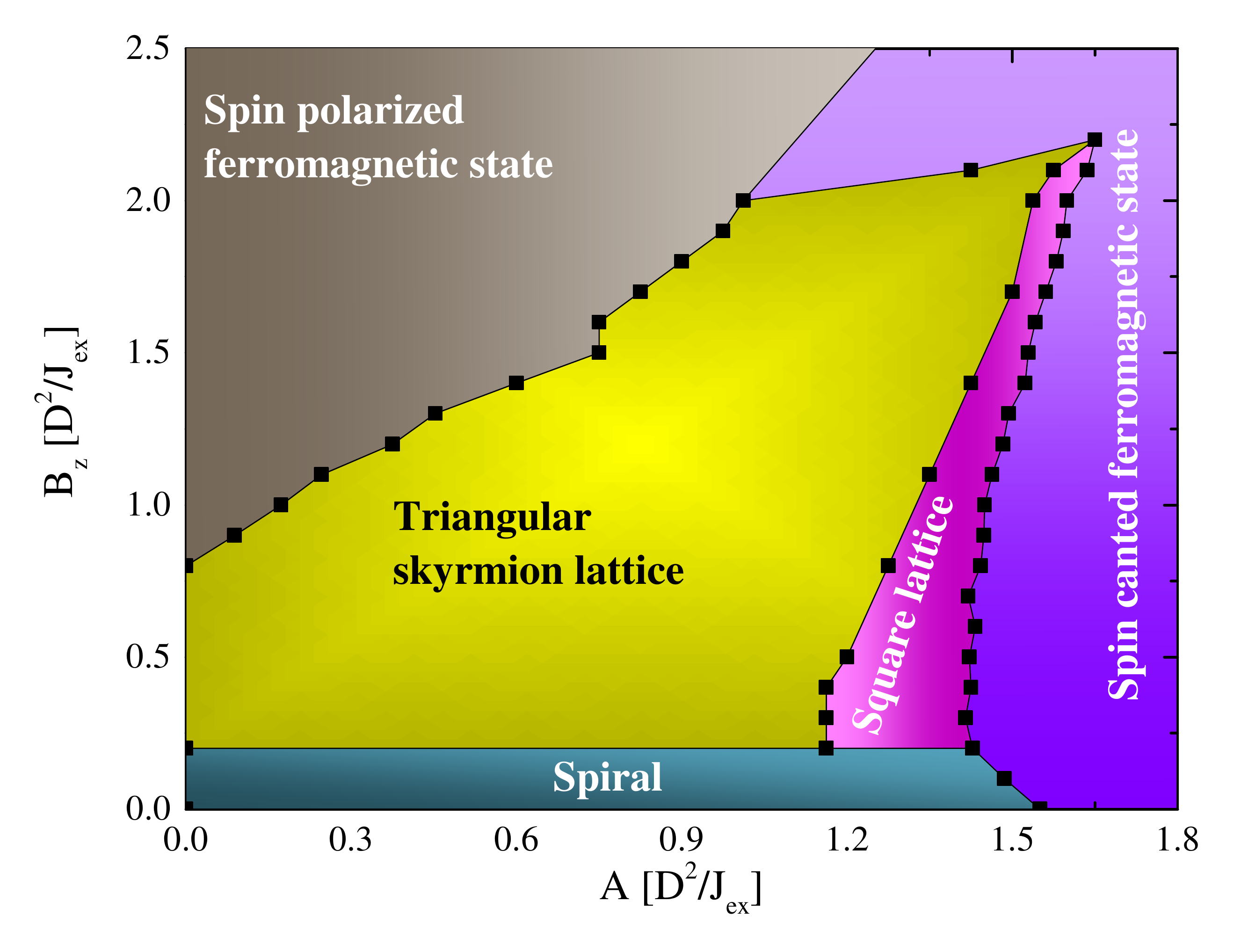, width=\columnwidth}
 \caption{(color online) Semi-quantitative $A$-$B_z$ phase diagram that results from comparing the energy obtained by annealing from the paramagnetic state to that of the uniform ferromagnetic state and the normal spiral state.} 
 \label{f6}
 \end{figure}

The calculated phase diagram is displayed in Fig.~\ref{f6}. Except for the second order transition between the two different spin polarized ferromagnetic states, the rest of the transitions are of the first order. Our annealing calculations fail to provide accurate phase boundaries  because the system gets trapped in metastable states. For this reason, the phase diagram depicted  in Fig. \ref{f6} must be regarded only as semi-quantitative. While this phase diagram has similarities with  the one obtained with a variational approach in Ref.~\onlinecite{Banerjee2014}, the variational phase diagram does not contain the square vortex lattice that is revealed by our unbiased numerical simulations. We note however that a tendency to form a square lattice was observed in the 2D square-cell calculation of Ref.~\onlinecite{Banerjee2014}, where the spin texture adopts a square shape and there are two peaks in the skyrmion topological charge density along the radial direction in the unit cell [see Fig.~4 (b) and (d) in Ref. \onlinecite{Banerjee2014}].

 The  anisotropy  required  to stabilize the square vortex lattice and the metastable meron states for typical  parameters of MnSi, $J_{\mathrm{ex}}\approx 3$ meV and  $D\approx 0.3$ meV~\cite{Zang11}, is  $A\approx 0.025$ meV. Our novel meron spin texture could be realized in MnSi ultrathin films because the easy-plane anisotropy increases significantly when the film thickness is reduced ~\cite{Karhu2012}. For MnSi, the lattice constant is $a=4.558\AA$ \cite{Janoschek2010} and the required anisotropy density is $A/a^3\approx 40\ \mathrm{KJ/m^3}$, which can be attained for films with thickness of $10$ nm according to the measurements from Ref.~\onlinecite{Karhu2012}.  Alternatively, spin anisotropy can be controlled by growing the chiral magnet films on  heavy element substrates, such as iridium ~\cite{Heinze2011,Romming2013}.  Note that the meron-antimeron spin texture is not  adiabatically connected to the skyrmion lattice because the total topological charge is different. Lorentz transmission electron microscopy can be used  to visualize the corresponding spin configuration. The skyrmion and meron configurations can also be imaged with a magnetic force microscope~\cite{Milde2013}. 

We emphasize that our phase diagram has been obtained for a single-layer system, i.e., for the strictly two-dimensional Hamiltonian ${\cal H}$ of Eq.~\eqref{eq1}.  For the multilayer extension of this Hamiltonian, the canted ferromagnetic state becomes a conical phase that turns out to be more stable than the skyrmion lattice phase. Consequently, a multi-layer system, like a thin film of finite thickness  $d\lesssim J_\mathrm{ex} a/D $, will exhibit a phase diagram like the one depicted 
in Fig.~\ref{f6} as long as the triangular skyrmion lattice remains stable relative to the conical phase due to the presence of an additional interaction 
(not included in ${\cal H}$) or due to the specific boundary conditions. For $d\gg J_\mathrm{ex} a/D$ the conical state is the most stable phase in the presence of easy-plane anisotropy.~\cite{Wilson2014}

\section{Conclusions}

In summary, our results reveal a structural transition between the triangular skyrmion lattice and a square vortex-antivortex lattice induced by increasing easy-plane anisotropy. Moreover, we find metastable  meron spin textures in the neighborhood  of the first order transition between the square vortex-antivortex lattice and the canted ferromagnet.
The stable crystal structures that appear as a function of increasing easy-plane anisotropy  are consistent with a subset of compact packings of disks of two different radii~\cite{Kennedy04}. Note that the triangular to square lattice transition has also been observed and discussed for skyrmions in the context of quantum Hall effect~\cite{PhysRevLett.76.4584, PhysRevLett.79.3998,PhysRevB.64.060104}.   The
metastable merons can be manipulated with external currents. This process can be reproduced  in real chiral magnets, such as MnSi, by growing thin films of progressively smaller thickness.

\acknowledgments 

The authors thank Achim Rosch, Weida Wu, Marc Janoschek, Charles Reichhardt and especially Mohit Randeria for helpful discussions. Computer resources for numerical calculations were supported by the Institutional Computing Program at LANL. This work was carried out under the auspices of the NNSA of the US DOE at LANL under Contract No. DE-AC52-06NA25396, and was supported by the US Department of Energy, Office of Basic Energy Sciences, Division of Materials Sciences and Engineering.


%

\end{document}